# Thermoelectric Properties of Intermetallic Semiconducting RuIn$_3$ and Metallic IrIn$_3$

N. Haldolaarachchige[1], W. A. Phelan[2], Y.M. Xiong[1], R. Jin[1], J. Y. Chan[2], S. Stadler[1] and D.P. Young[1*]

[1]*Department of Physics and Astronomy, Louisiana State University, Baton Rouge, Louisiana 70803, USA*

[2]*Department of Chemistry, Louisiana State University, Baton Rouge, Louisiana 70803 USA*

[*]Author whom correspondence should be addressed: dyoung@phys.lsu.edu

Low temperature (<400 K) thermoelectric properties of semiconducting RuIn$_3$ and metallic IrIn$_3$ are reported. RuIn$_3$ is a narrow band gap semiconductor with a large *n-type* Seebeck coefficient at room temperature $\left(S(290K) \approx -400 \ \mu\text{V/K}\right)$, but the thermoelectric Figure of merit $\left(ZT(290K) \approx 0.007\right)$ is small because of high electrical resistivity and thermal conductivity ($\kappa$(290 K) ~ 2.0 W/m K). IrIn$_3$ is a metal $\left(n(290K) \approx 10^{21} \text{cm}^{-3}\right)$ with low thermopower at room temperature $\left(S(290K) \approx -20 \ \mu\text{V/K}\right)$. Iridium substitution on the ruthenium site has a dramatic effect on transport properties, which leads to a large improvement in the power factor $\left(\dfrac{S^2}{\rho}(390\text{K}) \sim -207 \ \dfrac{\mu\text{W}}{\text{m}\cdot\text{K}^2}\right)$ and corresponding Figure of merit (*ZT*(380 K) = 0.053), improving the efficiency of the material by an over of magnitude.

## I.  INTRODUCTION

Improving the performance of thermoelectric materials is necessary if future refrigeration and power generation applications are to be realized. Since the discovery of Bi$_2$Te$_3$,[1, 2] thermoelectric materials have attracted attention because of high Seebeck coefficients, very low electrical resistivity, and high atomic weights.[3] Several interesting thermoelectric materials have been discovered more recently, such as skutterudites,[4] complex chalcogenides,[5] germanium based clathrates,[6] half-Heusler alloys,[7] a phonon glass–electron crystal system with exceptionally low thermal conductivity,[8] zintl phases,[9] and strongly correlated intermetallic semiconductors.[10-13] This last group of materials possesses a small hybridization gap at the Fermi level which is due to the mixing of the conduction band with narrow $d$ or $f$ bands. This is a very important characteristic for a large thermoelectric power ($S$).[14, 15] This is essential to having a high thermoelectric efficiency, which is quantified by the thermoelectric Figure of merit: $ZT = \left(S^2/\kappa_T \rho\right) T$, where $S$ is the Seebeck coefficient or thermopower, $T$ is the temperature, $\rho$ is the electrical resistivity, and $\kappa_T$ is the total thermal conductivity ($\kappa_T = \kappa_l + \kappa_e$, where $\kappa_l$ is the lattice or phonon contribution, and $\kappa_e$ is the electronic contribution).

RuIn$_3$ and IrIn$_3$ crystallize in the tetragonal FeGa$_3$ type (space group $P4_2/mnm$, No.136) structure.[16] RuIn$_3$ was first reported several decades ago,[17] however the electronic properties of RuIn$_3$ and IrIn$_3$ were first reported by Pottgen *et al.*,[18, 19] which showed poor metallic behavior. Later, Imai and Watanabe[16] observed semiconductor-like behavior for RuIn$_3$, which was confirmed by Bogdanove *et al.*[20] Electronic band structure calculations further show that RuIn$_3$ has a small band gap with a narrow peak in the density of states (DOS) close to the Fermi level. Such a feature often indicates a large thermopower, which is a necessary requirement for strong thermoelectric performance.[21]

Recent high pressure studies of this compound[22] show that the electronic structure is very sensitive to external pressure. Chemical pressure via doping may also lead to significant changes in the physical properties and enhance the thermoelectric efficiency, which has been observed in other systems, such as the skutterudites,[4] half-Heusler compounds[23] and other intermetallic semiconductors.[24] Indeed, chemical doping studies at the indium site[25-26] showed a large enhancement of the thermoelectric Figure of merit around 700 K. Takagiwa *et al.*[27] then performed experimental and theoretical studies of Ru(Ga/In)$_3$ and confirmed a considerably large thermoelectric Figure of merit at high temperatures.

The abovementioned measurements, however, focused only on the high temperature properties. Our main goal was to enhance the Figure of merit (*ZT*) near room temperature through chemical doping on the ruthenium site, as the valance band is dominated by Ru 4*d* states.[28] Such substitutions have proven successful in the past on several different thermoelectric materials[4, 7, 23], such as FeGa$_3$,[24] which showed significant changes in its physical properties and electronic structure with a large improvement of the thermoelectric Figure of merit close to room temperature. In this paper we present a detailed study of the low temperature electronic and thermoelectric properties of semiconducting RuIn$_3$ and metallic IrIn$_3$, such as electrical resistivity ($\rho$), Seebeck coefficient (*S*), thermal conductivity ($\kappa$), Hall-coefficient ($R_H$), calculated power factor ($S^2/\rho$), and thermoelectric Figure of merit (*ZT*).

## II. EXPERIMENTAL TECHNIQUES

Polycrystalline pure and chemically doped samples were prepared by heating stoichiometric amounts of starting materials in an rf-induction furnace under a partial pressure of ultra-high purity argon gas. The samples were then ground to fine powders and pressed into small pellets and annealed under vacuum at 800 °C for 48 hours to obtain a homogeneous sample. The crystal structure and phase purity of all the samples were investigated by powder X-ray diffraction using a Bruker AXS D8 advance diffractometer equipped with Cu Kα radiation. Electrical resistivity was measured by a standard four-probe method in a Quantum Design

Physical Property Measurement System (PPMS) using a bar-shaped sample $(1 \times 1 \times 3 \text{ mm}^3)$ from 380 K to 3 K. The Seebeck coefficient was measured in the PPMS from 380 K to 10 K using direct method[29] and comparative method with constantan standard. Thermal conductivity was measured using a longitudinal steady-state method in the PPMS from 380 K to 3 K. The room temperature Hall-coefficient was extracted from Hall resistivity data on a bar-shaped sample in the PPMS with a magnetic field to 9 Tesla.

## III. RESULTS AND DISCUSSIONS

Figure 1a shows the temperature dependent electrical resistivity of pure polycrystalline RuIn$_3$. It shows semiconductor-like behavior with a room temperature resistivity of 0.35 Ω cm and a carrier density of $1.8 \times 10^{18} \text{cm}^{-3}$, which agrees well with the previously reported values for single crystals.[20] Two regions are distinctively identified in the resistivity (region I: above 150 K: extrinsic response region, and region II: below 150 K: freeze out region).[22]

The temperature dependent Seebeck coefficient of pure polycrystalline RuIn$_3$ is shown in Figure 1b. At room temperature it has a large *n*-type Seebeck coefficient ($S$(290 K) = −419 µV/K). The magnitude of the thremopower (|$S$|) is weakly temperature dependent above 150 K, and rapidly decreasing toward zero below 150 K. The low temperature thermoelectric data have never been reported for pure RuIn$_3$, however this behavior is similar to what was recently reported for Sn- or Zn-doped RuIn$_3$ systems.[26] The negative sign of the Seebeck coefficient is consistent with the observed negative Hall-coefficient ($R_H$(290K)$_{RuIn3}$ = −3.34×10$^{-6}$ m$^3$/C) at room temperature, and agrees well with the literature.[20, 26] Electrons being the majority carrier are consistent with their higher mobility and smaller effective mass than the hole carriers (Bogdanove et al.[20], $m^* \sim 1.1\, m_0$, where $m_0$ is the mass of a free electron). A gradual decrement of the thermopower is due to the thermal excitation of electrons across the band gap, which is similar to the reported thermopower behavior of semiconducting $FeGa_3$.[13, 24]

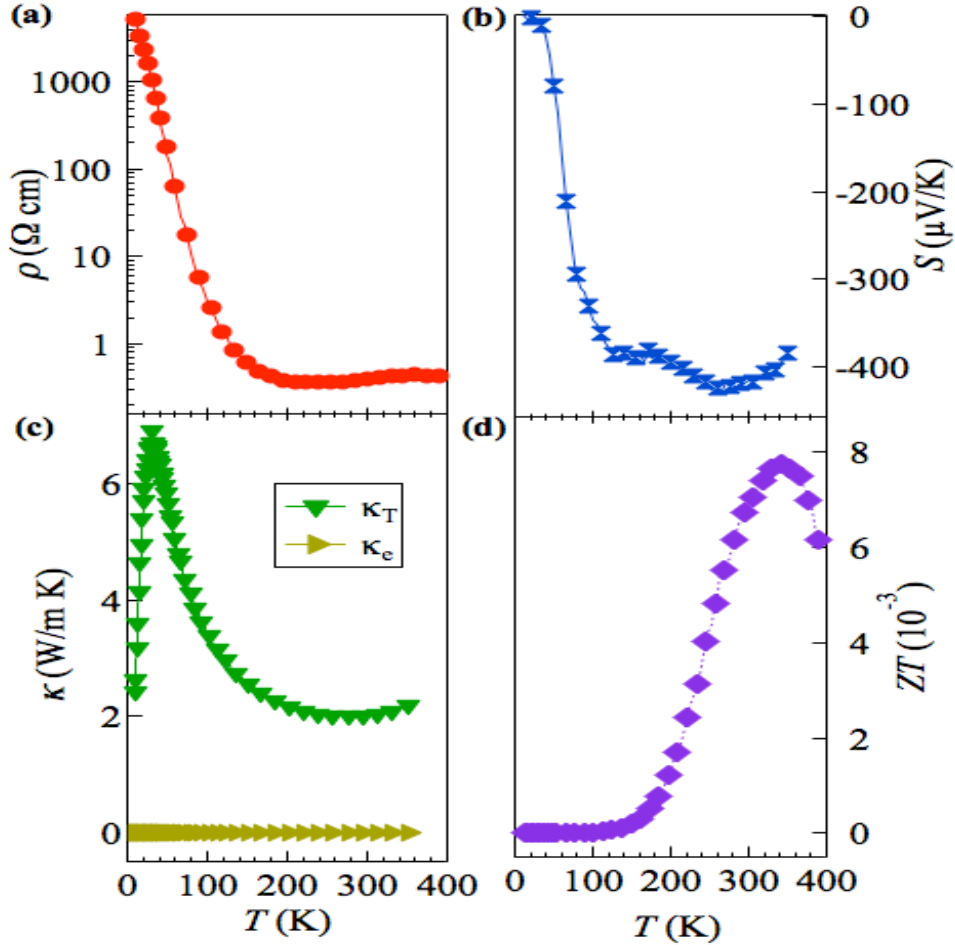

Figure 1    (Color Online) (a) Temperature dependent electrical resistivity, (b) Seebeck coefficient, (c) thermal conductivity, and (d) calculated thermoelectric Figure of merit of pure polycrystalline $RuIn_3$. Doted lines are guide to the eye.

Temperature dependent thermal conductivity (total: $\kappa_T$, and the electronic: $\kappa_e$) of pure polycrystalline $RuIn_3$ is shown in Figure 1c. The electronic thermal conductivity was estimated with the Wiedeman-Franz law ($\kappa_e = L_0 T/\rho$), where $L_0$ is the Lorentz number $\left(2.45\times10^{-8}\,\text{W}\Omega/\text{K}^2\right)$, $\rho$ is the electrical resistivity, and $T$ is the temperature. The calculated electronic thermal conductivity of RuIn$_3$ is almost negligible over the whole temperature range when compared with the total thermal conductivity, which is expected for intermetallic semiconducting compounds. A low temperature maximum in the total thermal conductivity is a characteristic feature of crystalline solids. However, the total thermal conductivity at room temperature ($\kappa_T$(290 K)= 2.0 W/m·K) is larger than that of the best thermoelectric materials.[3] The low temperature thermal conductivity of this compound has never been reported; however, the room temperature thermal conductivity is approximately 50% lower than that recently reported by Wagner et al.[25]

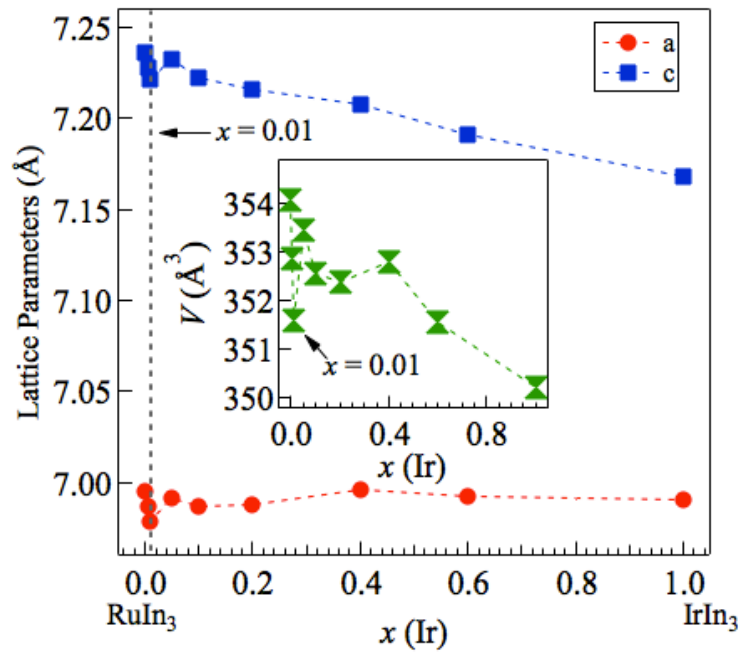

Figure 2     (Color Online) Calculated unit cell dimensions of pure and iridium doped polycrystalline RuIn$_3$ as a function of doping ($x$). Inset shows the unit cell volume as a function of $x$. Doted lines are guide to the eye.

The calculated thermoelectric Figure of merit of RuIn$_3$ ($ZT = S^2T/\rho\kappa$) is shown in Figure 1d. The Figure of merit at room-temperature is fairly low ($ZT$(290 K) = 0.007) because of the high thermal conductivity and high electrical resistivity. However, $ZT$(290 K) is slightly higher than that of some other intermetallic thermoelectric materials,[13] but much smaller than that of the best thermoelectrics.[3] The temperature dependent thermoelectric Figure of merit displays a maximum near room temperature.

Standard 2$\theta$ X-ray diffraction (XRD) patterns obtained from all the samples (pure and chemically-doped) matched the XRD pattern of RuIn$_3$. There were no impurity peaks of elemental or secondary phases detected in the XRD patterns. Figure 2 shows the calculated unit cell dimensions of pure polycrystalline RuIn$_3$ ($a$ = 6.995Å and $c$ = 7.236 Å), and the Ir-doped samples. The data for the pure samples agree well with the literature.[20] Ir substitution into polycrystalline RuIn$_3$ shows that both lattice parameters $a$ and $c$ gradually decrease with increasing doping level, and the behavior agrees well with Vegard's law, indicating a gradual shrinking of the unit cell volume from RuIn$_3$ ($V_{\text{RuIn3}}$ = 354.057 Å$^3$) to IrIn$_3$ ($V_{\text{IrIn3}}$ = 350.229 Å$^3$). Even though, overall, Vegard's law is obeyed, there is an exception at the 1% Ir doping level, where an unexpected sudden shrinking of the cell volume was observed. Interestingly, this concentration shows a large enhancement of its room temperature power factor and Figure of merit as will be shown below.

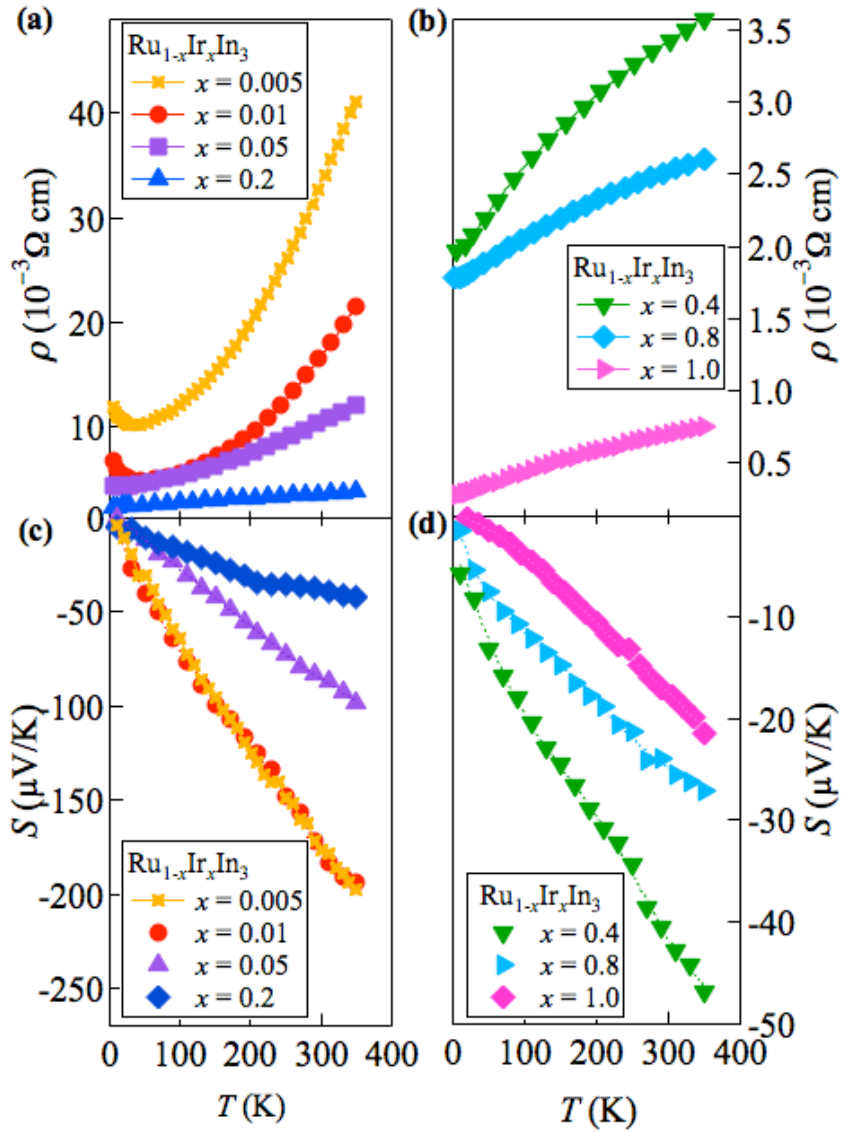

Figure 3 (Color Online) Temperature dependent electrical resistivity (**a** & **b**) and Seebeck coefficient (**c** & **d**) of chemically doped polycrystalline $RuIn_3$. Doted lines are guide to the eye.

Figures 3 (a) and (b) show the temperature dependent electrical resistivity of chemically doped polycrystalline $RuIn_3$. Electron doping shows a significant effect on the electrical resistivity. A small percentage (0.5%) of Ir substitution changes the semiconductor-like behavior of the pure compound into a metallic state with a lower electrical resistivity ($\rho$(290K))$_{Ru0.995Ir0.005In3}$ = 0.04 Ω-cm) and an increase in the carrier density $\left[n(290K) = 4.6 \times 10^{18} \text{ cm}^{-3}\right]$ at room temperature, which further confirms that the electronic structure of the pure compound can be effected even with a small amount of chemical doping.[22,25] Ir substitution for Ru at the 1% level decreases the room temperature resistivity by order of magnitude than that of the pure compound. Similar behavior was observed on 1.6% of Sn on In site, however 1.6% of Zn substitution amazingly reduced electrical resistivity by two orders of magnitude than the pure compound,[25] which is a promising result for enhancing the thermoelectric Figure of merit. Increasing the doping level of Ir (0.5% to 100%) results in metallic-like behavior over the whole measured temperature range and a lowering of the room temperature electrical resistivity all the way to pure IrIn$_3$ ($\rho$(290 K)$_{IrIn3}$ = 0.7 mΩ-cm) and an increase in the carrier density by two orders of magnitude $\left[n(290K) = 125 \times 10^{18} \text{ cm}^{-3}\right]$ higher than that of pure RuIn$_3$.

The temperature dependent thermopower of Ir doped RuIn$_3$ is shown in Figure 3 (c) and (d). A very small percentage of chemical substitution on the Ru site reduces the Seebeck coefficient ($S$(290 K)$_{Ru0.995Ir0.005In3}$ = -171 µV/K), which is consistent with the enhanced electrical conductivity and carrier density. The negative sign of the thermopower matches the negative Hall-coefficient ($R_H$(290K)$_{Ru0.995Ir0.005In3}$ = −1.37×10$^{-6}$ m$^3$/C) at room temperature. The percent change (55%) of the Seebeck coefficient of 1% Ir doping also agrees well with the recently reported Sn- and Zn-doped systems (75% for Sn and 62% for Zn).[25] Increasing doping levels of Ir (0.5% to 1%) show a slight improvement in thermopower ($S$(290 K)$_{Ru0.99Ir0.01In3}$ = −182 µV/K), and then 5% to 100% of doping decreases the room temperature thermopower toward the value of pure IrIn$_3$ ($S$(290 K)$_{IrIn3}$ ~ −20 µV/K).

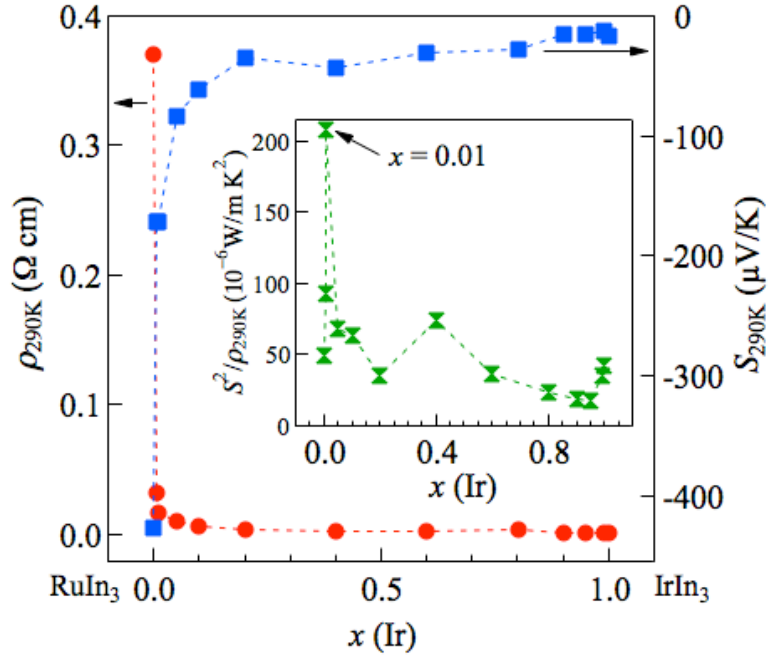

Figure 4　　(Color Online) Variation of room temperature resistivity, Seebeck coefficient and calculated power factor (inset) of iridium-doped polycrystalline RuIn$_3$. Doted lines are guide to the eye.

The variation of the room temperature electrical resistivity, Seebeck coefficient, and calculated power factor with increasing Ir substitution in RuIn$_3$ are shown in Figure 4. The room temperature electrical resistivity and thermopower decrease by a large percentage, even at very small doping levels. However, it can be clearly identified in the inset of Figure 4 that the room temperature power factor is maximized at 1% Ir substitution ($S^2/\rho$(290 K)$_{Ru0.99Ir0.01In3}$ = 209 $\mu$W/m·K$^2$), which is a factor of ~ 5 improvement over that of pure RuIn$_3$ and corresponds to the sudden collapse of the unit cell observed at the same doping level. Since RuIn$_3$ and IrIn$_3$ have about the same power factor (~ 47 $\mu$W/m·K$^2$) at room temperature, this is an important result showing the optimization of the thermoelectric properties at intermediate doping levels.

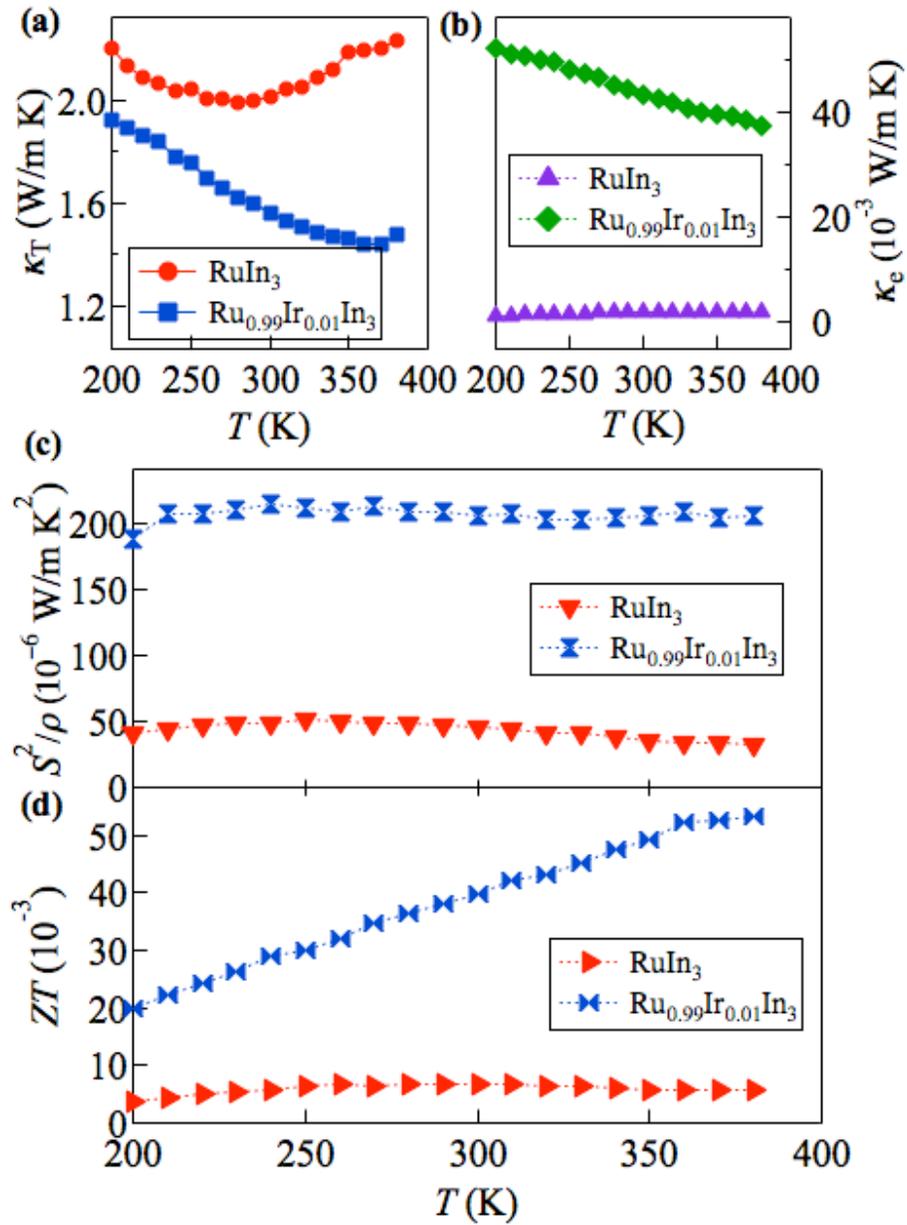

Figure 5  (Color Online) Temperature dependent thermal conductivity (**a** & **b**), calculated power factor (**c**) and Figure of merit (**d**) of pure and 1% of iridium doped polycrystalline $RuIn_3$. Doted lines are guide to the eye.

Figure 5 shows the temperature dependent thermal conductivity (total: $\kappa_T$, and electronic: $\kappa_e$), calculated power factor ($S^2/\rho$) and Figure or merit ($ZT$) of pure and 1% Ir-doped $RuIn_3$. The total thermal conductivity at room temperature slightly decreases (20%) with Ir doping over that of pure RuIn$_3$. Previous reports on Sn and Zn doping also indicated a significant decrease in the room temperature thermal conductivity.[25] Ru$_{0.99}$Ir$_{0.01}$In$_3$ shows a large increase in its electronic thermal conductivity (Figure 5b), which is expected with the enhanced carrier density in the system, but is still negligible when compared with the total thermal conductivity. The calculated power factor and the Figure of merit show a large enhancement over that of the pure compound over the whole measured range of temperature from 380 K to 200 K. The power factor ($S^2/\rho$(380 K)$_{Ru0.99Ir0.01In3}$ = 207 $\mu$W/m·K$^2$) increases by a factor of 5, and the Figure of merit ($ZT$(380 K) $_{Ru0.99Ir0.01In3}$ = 0.053) increases by a factor of 9 at 380 K over that of the pure compound for the 1% Ir-doped sample. This corresponds to slightly larger enhancement than recently reported for 1.6% Sn doping ($ZT \sim$ 0.04), but is smaller than 1.6% Zn doping ($ZT \sim$ 0.24).[25] The higher $ZT$ at 380 K for Zn substitution is due to the fact that a larger reduction of its electrical resistivity (2 orders of magnitude reduction than pure compound) takes place.

## IV. CONCLUSIONS

We have synthesized and characterized the low temperature thermoelectric properties of pure and chemically doped RuIn$_3$. Pure RuIn$_3$ is a semiconductor with a large *n*-type Seebeck coefficient, and a fairly large resistivity and thermal conductivity at room temperature, which leads to a small Figure of merit. A small amount of chemical substitution on the Ru site has a significant effect on the material's physical properties and electronic structure of the compound, which resulted in a substantial increase in the material's power factor with a slight lowering of the thermal conductivity. The highest power factor ($S^2/\rho$(380 K) = 207 $\mu$W/m-K$^2$) and corresponding Figure of merit ($ZT$(380 K) = 0.053) were observed for Ru$_{0.99}$Ir$_{0.01}$In$_3$.

## V.     ACKNOWLEDGEMENTS

DPY acknowledges the NSF grant no. DMR-1005764, JYC acknowledges the NSF-DMR 1063735, SS acknowledges the NSF grant DMR-0545728 and RJ acknowledges the NSF grant no. 1002622. NH acknowledges Dr. A. B. Karki for useful discussions.